\documentclass[aps,prb,twocolumn,superscriptaddress,groupedaddress]{revtex4}
 \usepackage{epsfig,amsopn}
\usepackage{graphicx}
\usepackage{amsmath,amssymb}
\usepackage{physics}
\usepackage{enumerate}
\usepackage{sidecap}
\usepackage[symbol]{footmisc}
\newcommand\bea{\begin{eqnarray}}
\newcommand\eea{\end{eqnarray}}
\newcommand\beq{\begin{equation}}
\newcommand\eeq{\end{equation}}

\newcommand{\bib}{\bibitem}
\def\nn{\nonumber}
\def\f{\frac}
\def\al{\alpha}

\def\si{\sigma}
\def\Do{\partial}

\def\dg{\dagger}
\def\la{\langle}

\def\ua{\uparrow}
\def\da{\downarrow}

\def\be{\beta}

\def\th{\theta}

\begin{document}

\title{Electron transport in  magnetic tunnel junctions
- a theoretical study of lattice and continuum models}
\author{Dhavala Suri}
\affiliation{ Department of Physics, Birla Institute of Technology and Science Pilani - K K Birla Goa Campus,
Zuarinagar, Goa 403 726, India.}
\affiliation{Francis Bitter Magnet Lab,
Massachusetts Institute of Technology, Cambridge, MA 02139, USA.}
\author{R. S. Patel}
\affiliation{Department of Physics, Birla Institute of Technology and Science Pilani - K K Birla Goa Campus,
Zuarinagar, Goa 403 726, India.}
\author{Abhiram Soori}\thanks{Corresponding author}
\affiliation{International Centre for Theoretical Sciences (TIFR), 151,
Shivakote, 
Bengaluru 560089, India.}
\affiliation{ Department of Physics, Indian Institute of Science, Bengaluru 560012, India.}
\affiliation{ School of Physics, University of Hyderabad, C. R. Rao Road, Gachibowli, Hyderabad-500046, India.}

\begin{abstract} 
Magnetic tunnel junctions comprising of an insulator sandwiched 
between two ferromagnetic films are the  simplest spintronic devices. 
Theoretically, these can be modeled by a metallic Hamiltonian in both
the lattice and the continuum with an addition of Zeeman field. We calculate 
conductance at arbitrary orientations of the easy axes of the two ferromagnets. When 
mapped, the lattice and the continuum models show a discrepancy in conductance 
in the limit of a large Zeeman field. We resolve the discrepancy by
modeling the continuum theory in an appropriate way.
\end{abstract}
\maketitle
\section{Introduction}

Magnetic tunnel junctions are basic building blocks of
 spintronic devices \cite{fabian}.  The experiments by
 Moodera~et.al~ and Miyazaki~et.al~ \cite{moodera95,miyazaki} (1995)
on magnetic tunnel junctions demonstrated a 
large tunnel magnetoresistance~(TMR) at room temperature 
and resulted in upsurge of activities in this field. Parkin et.al~\cite{parkin} 
used MgO as a tunnel barrier and enhanced the TMR to nearly 200~\%. A more recent 
experiment~\cite{ikeda}~(2008) reported a much higher value of TMR (604~\%) at 
room temperature.  
Theoretically, magnetic tunnel junctions have been modeled even recently 
with the objective of studying the electron transport across the 
junctions~\cite{pasanai15,sun,zheng,prl}. Other apsects of the continuum 
model we study in this work has been earlier addressed by 
Pasanai~\cite{pasanai15} in a one dimensional system. 

In experimental systems, the junction
can be tuned between ON and OFF states by an  externally applied magnetic 
field~\cite{moodera95,ikeda}. The ON state is when the spins in the two FM's are 
parallel and OFF state is when the two spins are anti-parallel. In the ON
state, the current between the two FM's due to an applied bias
is large and in the OFF state, it is small. However, theoretically
the two  FM's can be in a configuration that is more general than 
ON and OFF, where the spins
in the two ferromagnets are aligned at a relative angle $\th$ 
(with $0 \le \th \le \pi$). We shall use this as a
{\em theoretical tool} to understand the modeling of tunnel junctions. 
The experimental realizations generally correspond to the cases
$\th=0$ (ON) and $\th=\pi$ (OFF). 

In Sec.~\ref{sec-model}, we discuss the continuum and  lattice models 
that describe a ferromagnet, followed by an outline of a mapping from 
continuum to lattice model. In Sec.~\ref{sec-junction}, we discuss 
the modeling of the tunnel junction. In Sec.~\ref{sec-transport}, 
we calculate conductance. In Sec.~\ref{sec-discr}, we point to the 
discrepancy between the two models and resolve it. In Sec.~\ref{sec-summ}, 
we summarize and end with concluding remarks. 

\section{Models of ferromagnet}~\label{sec-model}
\subsection{Continuum model }
In continuum, an FM can be modeled by the Hamiltonian 
\beq H_c = (\hbar^2\vec k^2/2m-\mu_c+E_Z)\si_0 -E_Z\si_z, \label{h-cont} \eeq
where $\si_{i}$ for $i=0,x,y,z$ denote the Pauli spin matrices, the parameters 
$m,~\mu_c,~E_Z$ denote the effective mass of electrons, chemical potential
and the Zeeman energy respectively. The dispersions for up- and down-~spins
are respectively given by $E_{\uparrow}=\hbar^2\vec k^2/2m-\mu_c$ and 
$E_{\downarrow}=\hbar^2\vec k^2/2m-\mu_c+2 E_Z$.  By convention, Fermi energy 
is at zero and whenever a bias is applied, the bias window $[-eV_0, +eV_0]$ 
is placed at the Fermi energy. It is easy to see that the band-bottom 
for up- and down-~spins are at the energies $-\mu_c$ and ($2E_Z -\mu_c$) 
respectively. 

\begin{figure}[h]
\includegraphics[height=5cm]{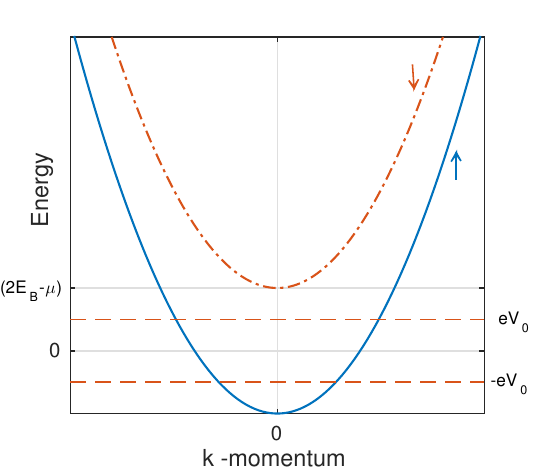}
\includegraphics[height=5cm]{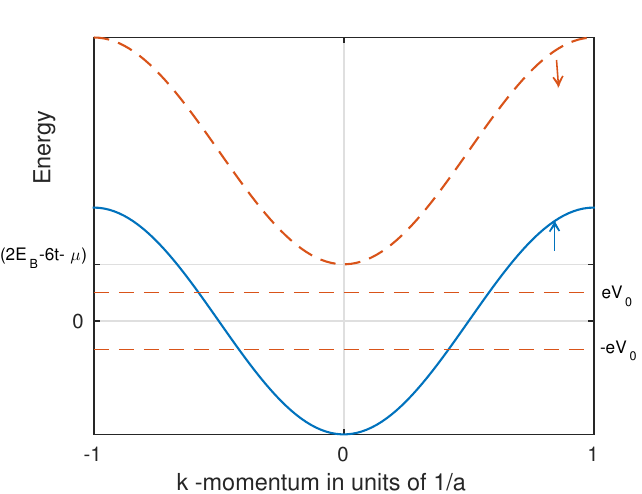}
\caption{Schematic of the dispersions for a ferromagnet modeled by 
continuum (top) and lattice (bottom) - Hamiltonians. $E_B=E_Z$, where 
$E_Z$ is the Zeean energy that can be found in
eq.~\eqref{h-cont} and eq.~\eqref{h-latt}. }~\label{fig01}
\end{figure}
\subsection{Lattice model}
The ferromagnet can also be modeled by a lattice Hamiltonian on a cubic lattice- 
\beq H_l = \sum_{\vec n, \vec e}^{} [ -t(c^{\dg}_{\vec n + \vec e}c_{\vec n} 
+ c^{\dg}_{\vec n}c_{\vec n + \vec e}) -(\mu_l -E_Z) c^{\dg}_{\vec n}c_{\vec n}
-E_Z c^{\dg}_{\vec n} \si_z c_{\vec n} ], \label{h-latt}\eeq 
where  $c_{\vec n}=[c_{\vec n,\uparrow}~, ~c_{\vec n,\downarrow}]^T$ and $c_{\vec n,\si}$ 
is the second quantized annihilation operator for the spin-$\si$ electron at site-$\vec n$, 
and $\vec e$ takes on unit vectors along $x,y,z$-directions.
The parameters $t,~\mu_l,~E_Z$
are the hopping strength, chemical potential and the Zeeman energy respectively for the 
lattice model. In certain limits, these parameters can be mapped on to the parameters
in the continuum model. The dispersion for the up-spin and the down-spin electrons in the 
lattice model take the form: $E_{\uparrow}=-2t[\cos(k_xa)+\cos(k_ya)+\cos(k_za)]-\mu_l$ 
and $E_{\downarrow}=-2t[\cos(k_xa)+\cos(k_ya)+\cos(k_za)]-\mu_l+2 E_Z$. The 
parameter $a$ is the lattice spacing.

\subsection{Mapping between the two models of ferromagnet}~\label{sec-mapping}
A schematic of the dispersions for the lattice and the continuum is shown in Fig.~\ref{fig01}.
We work in the range of parameters: (i) $0<eV_0<\min(\mu_c,2E_Z-\mu_c)$ in the continuum and 
(ii) $0<eV_0<\mu_l<6t$ and $eV_0<(2E_Z-6t-\mu_l)$ in the lattice. The dispersion relation
for a continuum model is isotropic while that for the lattice model is not isotropic except
very close to the band bottom. We now discuss on how to map from the continuum model to a lattice 
model. The lattice model is an effective model and may not depict the underlying lattice
structure of the material, but agrees with the continuum model at low energies.
Starting from a continuum model, we set the Zeeman energy in the
lattice model to be the same ($E_Z$). The condition $\mu_l=\mu_c-6t$ ensures that the band 
bottoms of the two models are aligned. The lattice dispersion has the same form as in the 
continuum in the limit when $k_xa,k_ya,k_za \ll \pi/2$, which can be seen by Taylor expanding
the lattice dispersion around $\vec k = \vec 0$ and keeping terms upto second 
order in $\vec ka$. This gives us: 
$ta^2=\hbar^2/2m$. We choose the lattice constant $a$ by demanding that in the bias window,
the binomial expansion of the lattice dispersion is a good approximation: 
$a\ll\hbar \pi/2\sqrt{2m(\mu_c+eV_0)}$. Once $a$ is chosen to satisfy the above condition, 
$t=\hbar^2/2ma^2$.

When $E_Z$ is large enough so that $(2E_Z-\mu_c) > eV_0$ [same as $(2E_Z-\mu_l-6t) > eV_0$], 
the down spin branch of the dispersion does not lie in the bias-window 
and the FM has 100\% polarization with only up-spin channel being occupied
at zero temperature. We call such a ferromagnet a {\sl ``pure~FM''}.
At a finite temperature, the down-spin channel has a small but finite occupation and the 
polarization is less than 100\% in such {\sl pure FM}'s. The exact value of polarization
at a given temperature depends on how large $E_Z$ is and decreases monotonically with $E_Z$. 
We also introduce the limit $E_Z\to\infty$, in which we call the ferromagnet a 
{\sl ``perfect~FM''} since the polarization even at a finite temperature remains at
100\%.

\begin{figure} 
\begin{center}
 \includegraphics[width=8cm]{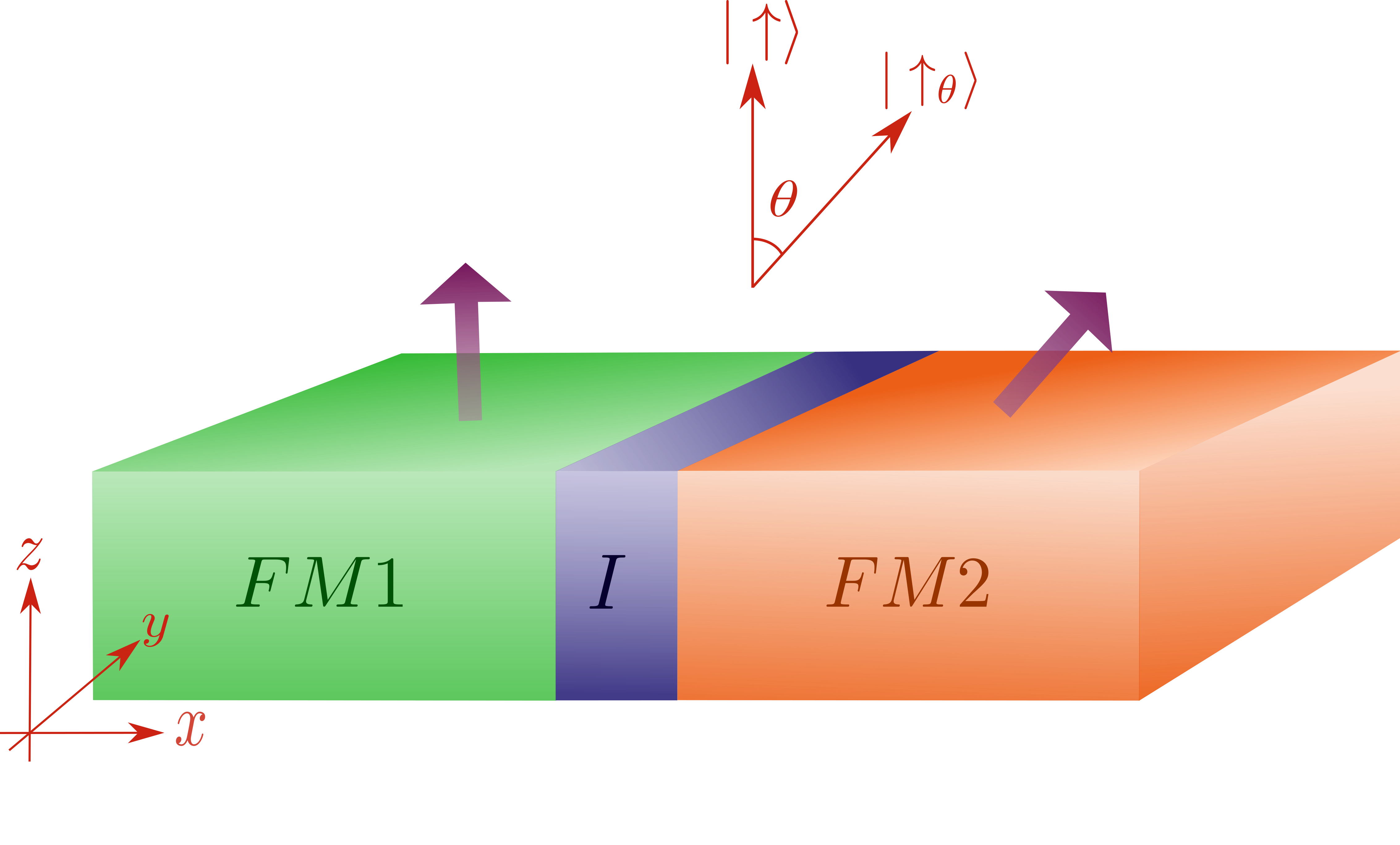}
\end{center}
\caption{Schematic of the setup under investigation. Two ferromagnets have their easy axes 
oriented at a relative angle $\theta$. A thin insulator is sandwiched between the two 
ferromagnets.}
\end{figure}

\section{Junction between two ferromagnets}~\label{sec-junction}
In continuum, a junction between two ferromagnets is typically described by respective 
Hamiltonians on either sides of $x=0$ with boundary condition~(BC) applied to the 
wavefunction at $x=0$. A junction between FM's pointing in two directions which are at 
angle~$\theta$ is modeled when the Hamiltonian on left is given by Eq.~(\ref{h-cont})
and the Hamiltonian on the right is given by  Eq.~(\ref{h-cont}), where $\si_z$ is 
replaced by $\si_{\th}=(\cos{\th}~\si_z+\sin{\th}~ \si_x)$. 
The wavefunctions on either sides of the junction are equal and their 
derivatives differ by a quantity proportional to the amplitude of the wavefunction at 
the junction-point: 
\bea \psi(x=0^+) &=&  \psi(x=0^-) \nn \\ 
\partial_x \psi|_{x=0^+} - \partial_x \psi|_{x=0^-} &=& q_0~\psi(x=0), ~\label{eq-bc}\eea
where $q_0$ parametrizes the transparency of the junction. The limits $q_0=0$ and 
$|q_0|\gg\sqrt{2 m \mu_c}/\hbar$ correspond to fully transparent and fully opaque junctions. 

In the lattice model, the junction is typically characterized by a hopping $t'$
between two sides of the lattices governed by different Hamiltonians and 
the BC does not appear. The information about the BC 
(of the continuum model) is carried by the hopping element $t'$ in the 
$\hat x$-direction. The Hamiltonian on the 
left $H_L$ is given by eq.~\eqref{h-latt} where $\vec n = (n_x,n_y,n_z)$ and 
$n_x,n_y,n_z$ takes integer values such that $n_x \le 0$. The Hamiltonian
on the right $H_R$ is given by eq.~\eqref{h-latt} where $\vec n = (n_x,n_y,n_z)$ and
$n_x,n_y,n_z$ take integer values such that $n_x \ge 1$, and  
$\si_z$ replaced by $\si_{\th}=(\cos{\th}~\si_z+\sin{\th}~ \si_x)$  like in the 
continuum. The full Hamiltonian is: 
\bea  H &=& H_L+H_R+H_T,~~~{\rm where} \nn \\ 
  && H_T=-t'\sum_{n_y,n_z} 
[c^{\dagger}_{(1,n_y,n_z)} c_{(0,n_y,n_z)} + h.c], ~\label{eq-hlatt}
\eea
and  $n_y,n_z$ run over all integers. 

We shall now specialize to the {\sl perfect ferromagnet} limit, where
$E_Z\to\infty$. Physically, this makes sense when the $E_Z$ for the 
ferromagnet is much larger than all other energy scales in the problem.
Also, this makes the calculations much simpler since one of the two 
spin channels tends to be absent. 

\section{Transport Calculations in continuum and lattice models}~\label{sec-transport}
We follow Landauer-B\"uttiker scattering approach~\cite{landau-butti,lb4,landau-butti2,landau-butti3,dattabook} 
to calculate conductance in continuum and lattice models. We write down
a wavefunction which has incident and scattered parts and solve for the 
scattering coefficients. Since this is a three dimensional system, there
are two angles of incidence and the total current at a given bias is 
calculated by integrating the currents over the full range of angles of 
incidence with appropriate factors. In the scattering theory calculation
for tunnel junctions, an electron is incident at the junction and 
scattering  amplitudes for scattering into different channels is calculated. 
Let $\alpha$ and $\be$ be the angles made by the incident electron 
having momentum $\vec k = (k_x,k_y,k_z)$ such that $\vec k= k(\cos{\alpha},
\sin{\alpha}\cos{\be},\sin{\alpha}\sin{\be})$. $\alpha$ is the 
angle made by the incident electron with $x$-axis, while $\be$ is the
angle made by the projection of the momentum $\vec k$ onto the $(y,z)$~plane
with $y$-axis. Due to translational invariance along $\hat y$- and $\hat z$-
directions, the momenta $k_y$ and $k_z$ are good quantum numbers.

In continuum, the wavefunction of an electron incident on the
tunnel junction at an energy $E$ from left lead will look like 
$e^{i(k_y y +k_z z)}\ket{\psi(x)}$ where,  
\bea  
\ket{\psi(x)} &=& (e^{i k_x x} + r_{E,\al}~ e^{-i k_x x})~\ket{\ua}
         ~~{\rm for}~~x<0,\nn \\
        &=& t_{E,\al}~ e^{i k_x x}~\ket{\ua_{\th}} 
        ~~{\rm for}~~x>0,
\eea
where $k_x=\sqrt{2m(\mu_c+E)}\cos{\al}/\hbar$ and the kets
denote the spinors: $\ket{\ua}= [1,~0]^T$, $\ket{\da}= [0,~1]^T$,
$\ket{\ua_{\th}}=[\cos{(\th/2)},~\sin{(\th/2)}]^T$ and $\ket{\da_{\th}} 
= [-\sin{(\th/2)},~\cos{(\th/2)}]^T$. Here, the wavevectors in the down-spin
channels ($\ket{\da}$ for $x<0$, and $\ket{\da_{~\th}}$ for $x>0$) are absent 
since we have taken the limit of {\sl pure ferromagnet}. 
To solve for the scattering amplitudes $r_{E,\al}$ and $t_{E,\al}$, 
we employ the boundary conditions discussed in Eq~\eqref{eq-bc}. It is easy to 
see from here that for any nonzero~$\th$, $t_E=0$ and hence the conductance
of the junction is zero. This comes as a surprise. 
At $\th=0$, the problem becomes that of a 
spinless tunnel junction the conductance of which is dictated by the barrier
strength $q_0$. 

\begin{figure}
\begin{center}
 \includegraphics[width=4.2cm]{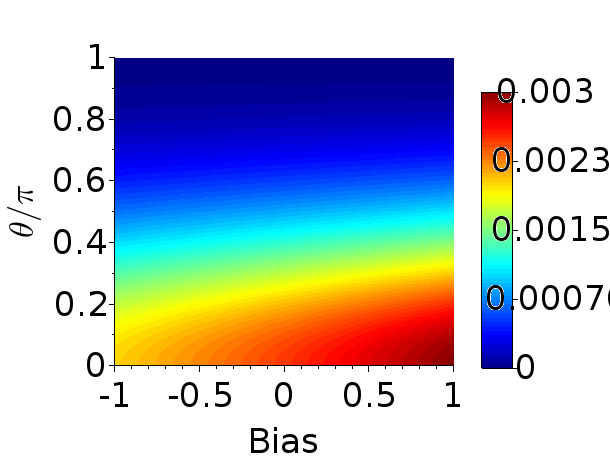} 
 \includegraphics[width=4.2cm]{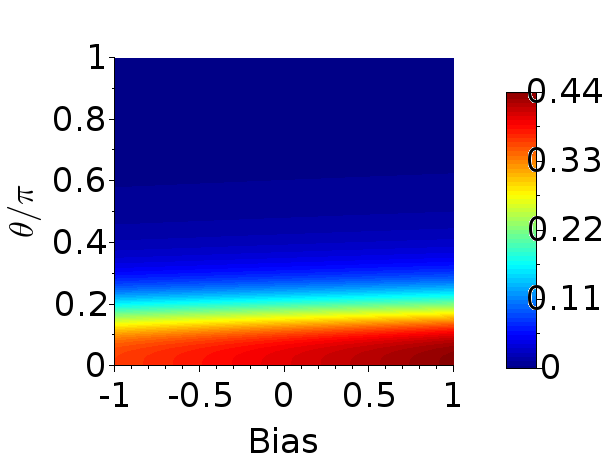}
 \end{center}
 \caption{Dependence of conductance in units 
 of $e^2/h$ on the bias and angle $\th$ 
 shown in contour plot for lattice model. Parameters: $t'=0.535t$ for left
 and $t'=t$ for right, $\mu_l=-5.994~t$ for both.}~\label{fig-latt}
\end{figure}

In the lattice model, the  wavefunction of an electron incident on the tunnel junction
at an energy $E$ from the left lead takes the form $\ket{\psi} = \sum_{n_x,n_y,n_z}
e^{i(n_yk_ya+n_zk_za)} \ket{\psi_{n_x}}\ket{n_x,n_y,n_z}$,
where $\ket{\psi_{n_x}}= \psi_{n_x} \Theta[-n_x+1] 
\ket{\ua} + \psi_{n_x} \Theta[n_x] \ket{\ua_{\th}}$, where $\Theta[n_x]$ is discrete Heaviside
step function and the kets retain the identity assigned in previous paragraph. This
just states that electrons can point only along the easy axis and the easy axes of
electrons on either sides of the junction differ by an angle $\th$. Further, the 
wavefunction takes the form: 
\bea \psi_{n_x} &=& (e^{i n_x k_x a} + r_{E,\al}~ e^{-i  n_x k_x a}),
         ~~~{\rm for}~~n_x \le 0,\nn \\
        &=& ~~t_{E,\al}~ e^{i n_x k_x a},~ 
        ~~~{\rm for}~~n_x\ge 1,
\eea
where $k_x=\sqrt{2m(\mu_l+6t+E)}\cos{\al}/\hbar$. We choose $\mu_l$ 
such that the lattice dispersion can be approximated to be a quadratic
dispersion near the band bottom, as discussed in section~\ref{sec-mapping}.
Note that $\mu_l$ so chosen is negative. The equation connecting 
the wavefunctions on either sides of the junction obtained from lattice 
Hamiltonian eq.~\eqref{eq-hlatt}, reduces to the following equations
in the limit of {\sl perfect ferromagnet}:
\bea  
E_x~\psi_0 &=& -t \psi_{-1}-t' \cos{(\th/2)}\psi_{1} \nn \\
E_x~\psi_1 &=& -t'\cos{(\th/2)} \psi_{0} -t \psi_{2}, ~\label{eq-hop}
\eea
where $E_x=(E+\mu_l+6t)\cos^2{\al}-2t$. 
The term in eq.~\eqref{eq-hop}
proportional to $t'$ can be understood as follows- 
since an $\ket{\ua}$~electron on site $n_x=0$ does not flip spin 
while hopping on to $n_x=1$, where the eigenspinor is $\ket{\ua_{\th}}$, 
the overlap is $\bra{\ua}\ket{\ua_{\th}}=\cos{(\th/2)}$. Solving for the 
scattering amplitudes using eq.~\eqref{eq-hop}, we get 
\beq t_{E,\al} = \frac{-2 i \sin{(k_x a)} ~t ~t' \cos{(\th/2)}}{[t^2 ~e^{-ik_x a}
-t'^{~2} ~e^{ik_xa} \cos^2{(\th/2)}]}, \eeq
and the differential conductance is given by 
\bea
G &=& G_{0}(E)
\int_0^{\pi/2} d\al \sin{\al} \cos{\al}~|t_{E,\al}|^2, \nn \\ 
&&{\rm where~~} G_0(E) = \f{e^2}{h} \f{mA (\mu_c+E)}{\pi h^2}, ~\label{eq-cond}
\eea
and $A$ is the area of cross section of the tunnel junction. The 
energy dependent factor $(\mu_c+E)$ in the conductance reflects
the density of states in three dimensional ferromagnet. In Fig.~\ref{fig-latt}, 
we  show the result for differential conductance as a function of the angle
$\th$ and the bias.

\begin{figure}
\begin{center}
  \includegraphics[width=4.2cm]{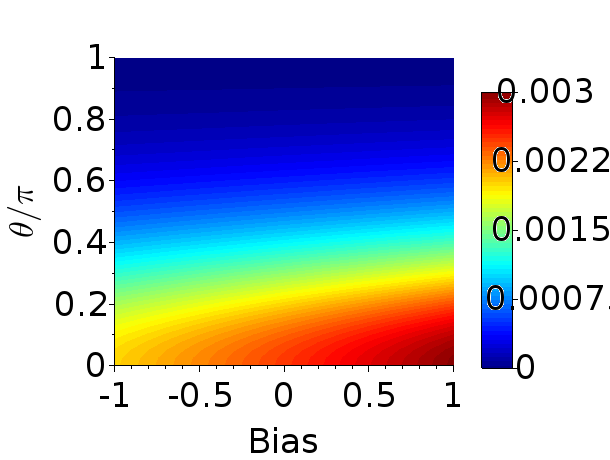} 
 \includegraphics[width=4.2cm]{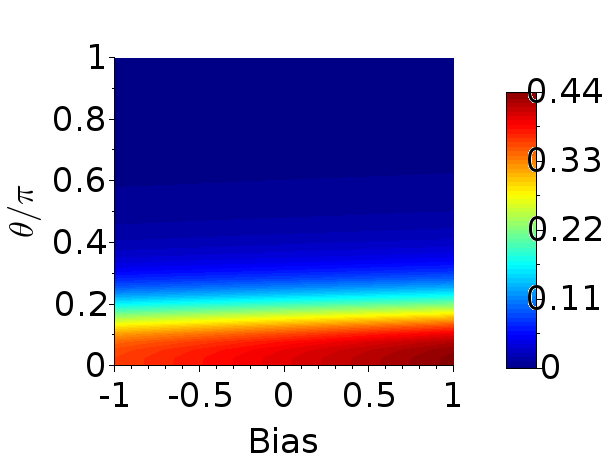}
 \end{center}
 \caption{Dependence of conductance in units 
 of $e^2/h$ on the bias and angle $\th$ 
 shown in contour plot for continuum model.
 Parameter $V_b=2000$ for left
 and $V_b=0$ for right. Other parameters have been 
 mapped from lattice to the continuum as discussed 
 in sec.~\ref{sec-mapping} : $m=10$, $\mu_c=10$, and 
 $a=0.0052952$.}~\label{fig-cont}
\end{figure}

\section{Discrepancy and its resolution}~\label{sec-discr}
In the last section~(sec.~\ref{sec-transport}) we demonstrated that
the lattice and the continuum models of the junction give highly distinct 
results for conductance of the magnetic tunnel junction made from 
{\sl perfect FM}'s. In the lattice model, the conductance is 
exactly zero only when $t'\cos{(\th/2)}=0$, while in the continuum 
model, the conductance is zero whenever $\th \neq 0$. At this point, 
the results from lattice model look more reasonable. To resolve the
puzzle, we now take lattice model as the starting point of analysis.
The physical meaning of the hopping term $H_T$ in the Hamiltonian
[eq.~\eqref{eq-hlatt}] is that the atomic wavefunctions on the two
lattice sites overlap and 
the overlap is proportional to $t'$. In contrast, the 
wavefunctions on the two sides of the junction in the continuum model 
cannot overlap with each other whenever $\th \neq 0$ since the limit 
$E_Z\to\infty$ does not allow any nonzero spin-component in a direction 
different from the easy axis. However, a region of finite length can be 
introduced between the two ferromagnets in the continuum model where 
both spin channels are allowed and the two wavefunctions can
overlap. We show that introduction of such a region resolves the
disagreement between the two models. The region in between can be
modeled to have a length~$a$ and a barrier $V_b$ that reflects the 
hopping element $t'$. The barrier $V_b$ in the region $0<x<a$ is 
isotropic in the spin space and the boundary conditions at $x=0$ and 
$x=a$ are continuity of wavefunction $\psi$ and its derivative $\Do_x\psi$.
The limits $t'=t$ and $t'\to 0$ correspond to 
$V_b=0$ and $V_b\to\infty$ respectively.
 
 The wavefunction of an electron incident on the
tunnel junction at an energy $E$ from left lead will look like 
$e^{i(k_y y +k_z z)}\ket{\psi(x)}$ where,  
\bea  
\ket{\psi(x)} &=& (e^{i k_x x} + r_{E,\al}~ e^{-i k_x x})~\ket{\ua}
         ~~{\rm for}~~x<0,\nn \\
         &=& \sum_{\si=\ua,\da; ~\nu=+,-} s_{\nu,\si} e^{i \nu k'_x x}
         \ket{\si}~~{\rm for}~~0<x<a,\nn \\ 
        &=& t_{E,\al}~ e^{i k_x x}~\ket{\ua_{\th}} 
        ~~{\rm for}~~x>a.
\eea
Continuity of $\ket{\psi(x)}$ at $x=0,a$ gives four equations while 
continuity of $\la \ua |\Do_x\ket{\psi(x)}$ at $x=0$ and continuity
of $\la \ua_{\th}|\Do_x\ket{\psi(x)}$ at $x=a$ totally give six equations
to be solved for six scattering amplitudes. We are interested in
$t_{E,\al}$, since conductance can be calculated from it using
eq.~\eqref{eq-cond}. We numerically compute $t_{E,\al}$ for each $\al$, 
and integrate over $\al$ as shown in eq.~\eqref{eq-cond} to get conductance
for a given bias $E=eV$. The result for continuum model with an
 insulating layer in between the two FM's is presented in
Fig.~\ref{fig-cont}. The parameters for the two models have been 
chosen such that the final result looks the same and we have demonstrated 
this in the fully transparent and weakly transmitting limits. The results
show a remarkable similarity both qualitatively and quantitatively.

 \section{Summary and discussion}~\label{sec-summ}
 To summarize, we started with models of ferromagnets and 
 using these as building blocks, studied magnetic tunnel junctions in both
 continuum and lattice models. In the limit of perfect FM, the 
 discrepancy between the continuum and the lattice models first came as a
 surprise. However, the results from the lattice model calculations appeared
 more convincing. We then resolved the discrepancy by appropriate modeling of the 
 system in continuum model. In other words, given a magnetic tunnel junction 
 described by a lattice model, we found the corresponding continuum model. 
 
 The central result of our work is that  magnetic 
 tunnel junctions in the continuum model comprise of a nonmagnetic insulating
 layer sandwiched between the two ferromagnets. Absence of such a 
 nonmagnetic region gives rise to unphysical results. However, in a lattice 
 model, there is no need for a nonmagnetic layer  for magnetic tunnel junction. 
 
 In this work we have primarily focused on perfect ferromagnets, where no modes 
 in the direction other than spin easy axis is allowed. It will be interesting to
 investigate the lattice and continuum models in the pure ferromagnets where the 
 wavefunction in the direction opposite to that of spin easy axis decays 
 exponentially away from the junction. In junctions comprising of pure FM's, 
 the  non-magnetic region may not be required in modeling since the wavefunctions
 opposite to spin easy axis can overlap in the region close to the junction.
 Further, connection to experimental  systems is another future direction. 
 
%

\section*{Acknowledgements}
 We thank Diptiman Sen for stimulating discussions. DS thanks DST-INSPIRE, Govt. of India
 for PhD fellowship (No. DST/INSPIRE Fellowship/2013/742). AS thanks DST under grant number DSTO1597 and
 DST-INSPIRE Faculty award scheme under award number IFA17-PH190.


\end{document}